\documentclass{article}
\usepackage{authblk}
\usepackage[a4paper,left=3cm,right=3cm,top=3cm,bottom=3cm]{geometry}
\RequirePackage{amsthm,amsmath,amsfonts,amssymb}
\RequirePackage[numbers]{natbib}
\RequirePackage[colorlinks,citecolor=blue,urlcolor=blue]{hyperref}
\RequirePackage{graphicx}

\newtheorem{theorem}{Theorem}[section]

\newtheorem{corollary}[theorem]{Corollary}
\theoremstyle{remark}
\newtheorem{definition}[theorem]{Definition}

\newcommand{\E}{\mathbb{E}}
\newcommand{\+}[1]{\mathbf{#1}}
\renewcommand{\O}{\mathbb{O}}
\renewcommand{\P}{\mathbb{P}}
\newcommand{\Ipq}{\+I_{p,q}}
\newcommand{\R}{\mathbb{R}}
\renewcommand{\tilde}{\widetilde}

\newcommand{\norm}[1]{\|#1\|}
\newcommand{\tti}[1]{\|#1\|_{2\to\infty}}
\newcommand{\infnorm}[1]{\|#1\|_{\infty}}

\newcommand{\Lsym}{\+L_{\text{sym}}}
\newcommand{\Lrw}{\+L_{\text{rw}}}

\DeclareMathOperator{\diag}{diag}
\usepackage{float}
\usepackage{placeins}
\usepackage{afterpage}
\usepackage{tikz}
\usetikzlibrary{shapes}  
\usepackage{algorithm}
\usepackage[noend]{algpseudocode}

\author{Alexander Modell}
\author{Patrick Rubin-Delanchy}
\affil{University of Bristol, U.K.}
\date{}
\title{Spectral clustering under degree heterogeneity: a case for the random walk Laplacian}

\begin{document}
\maketitle

\begin{abstract}
  This paper shows that graph spectral embedding using the random walk Laplacian produces vector representations which are completely corrected for node degree. Under a generalised random dot product graph, the embedding provides uniformly consistent estimates of degree-corrected latent positions, with asymptotically Gaussian error. In the special case of a degree-corrected stochastic block model, the embedding concentrates about $K$ distinct points, representing communities. These can be recovered perfectly, asymptotically, through a subsequent clustering step, without spherical projection, as commonly required by algorithms based on the adjacency or normalised, symmetric Laplacian matrices. While the estimand does not depend on degree, the asymptotic variance of its estimate does --- higher degree nodes are embedded more accurately than lower degree nodes. Our central limit theorem therefore suggests fitting a weighted Gaussian mixture model as the subsequent clustering step, for which we provide an expectation-maximisation algorithm.
\end{abstract}

\section{Introduction}
The task of finding communities in networks is ubiquitous in modern statistics. Spectral clustering is a popular algorithm which partitions the nodes based on the eigenvectors of a matrix representation of the graph \citep{ng2001spectral,von2007tutorial,shi2000normalized}. Together, these produce a low-dimensional vector representation of each node, known as a graph embedding, which is then input to a standard clustering algorithm. The popularity of spectral clustering can be largely attributed to its computational tractability (it is fast, even for very large graphs \citep{baglama2005augmented}) and theoretical guarantees. Graph embeddings, without the subsequent clustering step, are of independent interest, as they serve as the foundation of many other forms of graph inference \citep{belkin2003laplacian,coifman2005geometric,tang2013universally,fishkind2015vertex,tang2017semiparametric,tang2017nonparametric}. 

Typically, the matrix used in the embedding step is either the adjacency matrix or one of two related matrices --- the normalised, symmetric Laplacian or the random walk Laplacian. In a highly cited tutorial on spectral clustering \citep{von2007tutorial}, the random walk Laplacian is advocated for over the symmetric Laplacian on grounds of favourable graph conductance properties, yet, beyond this, its merits are relatively understudied. By contrast, a wealth of literature has emerged on the statistical properties of graph embeddings obtained from the adjacency and symmetric Laplacian matrices. The purpose of this paper is to fill this gap and provide a principled, statistical interpretation of graph embeddings obtained from the random walk Laplacian. We demonstrate that, in a sense to be made precise, spectral embedding using the random walk Laplacian produces vector representations which are completely corrected for node degree, making it a favourable choice for graph inference in the presence of degree heterogeneity.

The stochastic block model \citep{holland1983stochastic} is a canonical statistical model for the study of communities in networks. Each node of the graph is assigned to one of $K$ communities and edges between nodes occur independently with probabilities depending only on their community memberships. In influential work, Rohe et al. \citep{rohe2011spectral} showed that spectral clustering via the symmetric Laplacian produces consistent estimates of those communities (see also \citep{mcsherry2001spectral}). Since then, a vast literature has emerged on the statistical properties of spectral clustering under the stochastic block model \citep{sussman2012consistent,fishkind2013consistent,lyzinski2014perfect,sarkar2015role,lei2015consistency,lyzinski2016community,athreya2017statistical,abbe2017community}.

The stochastic block model is an example of a latent position model \citep{hoff2002latent}, in which each node, $i$, is represented by a low-dimensional vector, $X_i$, and edges occur, independently of each other, with probabilities given by some kernel function, $f(\cdot,\cdot)$, of the relevant vector representations. In the stochastic block model, each node is represented by one of $K$ distinct points, $v_1,\ldots,v_K$, corresponding to the $K$ communities.  A useful kernel function to consider in the context of spectral clustering is the indefinite dot product, $f(x,y) = x^\top \+I_{p,q} y$, where $\+I_{p,q}$ is the diagonal matrix of $p$ ones followed by $q$ minus-ones. Under this model, known as the generalised random dot product graph \citep{young2007random,nickel2008random,athreya2017statistical,rubin2017statistical}, the embedding obtained from the scaled eigenvectors of the adjacency matrix can be seen to be estimating  $X_1,\ldots, X_n$ \citep{sussman2013consistent}. Similarly, the embedding obtained from the symmetric Laplacian can be seen to be estimating $X_1/\sqrt{t_1},\ldots,X_n/\sqrt{t_n}$, where $t_1,\ldots, t_n$ are the expected degrees of each node \citep{tang2018limit}. In either case, it has been shown that the error of each estimate is asymptotically Gaussian \citep{athreya2016limit,rubin2017statistical,tang2018limit}, motivating the recommendation to fit a $K$-component Gaussian mixture model, in the subsequent clustering step, to recover the communities of a stochastic block model.

While the stochastic block model is an appealing and analytically tractable model for studying communities in networks, its usefulness in practice is disputed. This is frequently put down to the fact that nodes within the same community are required to have the same expected degree, a property which is rarely observed in real-world networks. Instead, real-world networks typically have degree distributions which are highly heterogeneous \citep{albert2002statistical,broido2019scale}.

To remedy this, the degree-corrected stochastic block model \citep{karrer2011stochastic} generalises the stochastic block model by introducing node specific weights  $w_1,\ldots,w_n$ (scalars), which describe the `activeness' or `popularity' of each node. The probability of observing an edge between nodes $i$ and $j$ is given by the relevant inter-community probability, multiplied by the product of the nodes' parameters, $w_i w_j$, allowing a node's expected degree to be independent of its community. This is said to provide a more realistic model for community-structured graphs while remaining analytically tractable.

When represented as a generalised random dot product graph, a node $i$ in community $k$ is represented as $X_i = w_i v_k$ and all the points corresponding to this community lie on a `ray', a line through the origin.

To estimate the communities of a degree-corrected stochastic block model, it is not possible to directly cluster the points obtained from spectral embedding the adjacency or symmetric Laplacian matrices: for nodes $i$ and $j$ in the same community we have neither $X_i = X_j$ nor $X_i/\sqrt{t_i} = X_j/\sqrt{t_j}$. The standard adjustment, introduced by Ng et al. \citep{ng2001spectral}, and employed extensively thereafter \citep{qin2013regularized,lyzinski2014perfect,lei2015consistency,passino2020spectral}, is to project the spectral embeddings onto the unit sphere and subsequently perform clustering on these points. This projection step is intended to remove the ancillary effect of degree heterogeneity on the embedding.

The subject of this paper is spectral clustering via an alternative matrix representation --- the random walk Laplacian --- the transition matrix of a random walk on the graph. We will demonstrate that the embedding obtained this way can be viewed as estimating $X_1/t_1, \ldots, X_n/t_n$, which we will herein refer to as the {\it degree-corrected latent positions}. Under the degree-corrected stochastic block model, nodes of the same community have the same degree-corrected latent position, and each lies in one of $K$ distinct places, corresponding to the $K$ communities, much like the standard latent positions of a standard stochastic block model.

One way or another, to correct a $d$-dimensional spectral embedding for node degree, a method will typically seek a projection of the nodes onto a $d-1$-dimensional submanifold. This manifold is often, but not always \citep{jin2015fast}, a sphere \citep{ng2001spectral,qin2013regularized,lyzinski2014perfect,lei2015consistency,passino2020spectral}. However, in geometry, the usual way of representing the space of lines through the origin is with a hyperplane in which each point represents the line going through it, known as projective space \citep{lee2013smooth}. This is the representation we get by considering $X_1/t_1, \ldots, X_n/t_n$ (points on a hyperplane), and it is also the representation implicit in random walk spectral embedding (points on a different hyperplane). This has the significant practical advantage of being reducible to $\R^{d-1}$ without distortion.

In this way, a practitioner need only know what our theoretical results say about those $\R^{d-1}$ representations: as the number of nodes in the graph goes to infinity, the random walk spectral embedding provides uniformly consistent estimates of the degree-corrected latent positions, with asymptotically Gaussian error (up to identifiability and assuming expected degrees grow polylogarithmically).

Under a degree-corrected stochastic block model, this means that the estimates converge to $K$ distinct places, corresponding to the $K$ communities, allowing asymptotically perfect clustering. Additionally, in a sparse regime, our central limit theorem shows that the scale of the error is inversely proportional to the node's expected degree: higher degree means higher precision. We therefore propose to fit a weighted Gaussian mixture model to the random walk spectral embedding, which gives higher degree nodes more influence, and provide an expectation-maximisation algorithm to do so.

In our simulation study, we will compare this approach to alternatives which follow the steps: spectral embedding, degree-correction, clustering \citep{ng2001spectral,qin2013regularized,lyzinski2014perfect,lei2015consistency,jin2015fast,passino2020spectral}. There are many other approaches to performing community detection under the degree-corrected stochastic block model. Of the spectral variety, Chaudhuri et al. \citep{chaudhuri2012spectral} propose a 7 step procedure, involving a random graph split, spectral embedding of regularised forms of the random walk or symmetric Laplacian, and sequential clustering. Coja-Oghlan and Lanka \citep{coja2010finding} and Gulikers et al. \citep{gulikers2017spectral} propose a sequential clustering procedure based on embedding an alternative matrix representation, which could be described as a doubly normalised symmetric Laplacian. Outside of the domain of spectral clustering, a wealth of alternative methods have been proposed \citep{zhao2012consistency,amini2013pseudo,peng2016bayesian,chen2018convexified,gao2018community}.

The remainder of this article is organised as follows. In Section~\ref{sec:rwl}, we define the random walk Laplacian, discuss some of its properties and define spectral embedding via the random walk Laplacian. In Section~\ref{sec:models}, we define the generalised random dot product graph and a special case, the degree-corrected stochastic block model. Section~\ref{sec:theory} presents asymptotic theory which supports the interpretation that random walk spectral embedding estimates the degree-corrected latent positions of a generalised random dot product graph, including the degree-corrected stochastic block model as a special case. In Section~\ref{sec:spec_clust}, we give the algorithmic details of our procedure for estimating community structure under a degree-corrected stochastic block model and compare it to existing methods
 in a simulation study. Section~\ref{sec:real-data} provides an example application of estimating community structure in a character network. Finally, Section~\ref{sec:conclusion} concludes.

\section{The random walk Laplacian}
\label{sec:rwl}
Given a simple, undirected, connected graph with (symmetric) adjacency matrix $\+A \in \{0,1\}^{n \times n}$, with a one in position $i,j$ if there is an edge between nodes $i$ and $j$ and a zero otherwise, the symmetric Laplacian, $\Lsym$, and random walk Laplacian, $\Lrw$, are defined as
\begin{equation*}
  \Lsym := \+D^{-1/2}\+A\+D^{-1/2},\qquad  \Lrw := \+D^{-1}\+A,
\end{equation*}
where $\+D \in \R^{n \times n}$ is the diagonal degree matrix with entries $\+D_{ii} = \sum_j \+A_{ij}$. The reader may be more familiar with the definitions $\+I - \Lsym$ and $\+I - \Lrw$ \citep{von2007tutorial}. Both definitions share the same eigenvectors, so for our purposes they are equivalent.

The random walk Laplacian defines the transition matrix of a random walk on the graph and is closely related to the symmetric Laplacian. The following are some important properties relating the two matrices (see \citep{chung1997spectral} for a comprehensive review).
\begin{enumerate}
  \item If $\lambda$ is an eigenvalue of $\Lrw$ with corresponding eigenvector $u$, then $\lambda \in [-1,1]$ and $u$ is real-valued.
  \item The all-one vector $\mathbf{1}$ is an eigenvector of $\Lrw$ with eigenvalue 1.
  \item $\lambda$ is an eigenvalue of $\Lrw$ with eigenvector $u$ if and only if $\lambda$ is an eigenvalue of $\Lsym$ with eigenvector $\+D^{1/2}u$.
\end{enumerate}
We note that these properties hold when $\+A$ is replaced by any non-negative matrix.

We consider the following spectral embedding of $\Lrw$ into $\R^{d-1}$. The notation $|\cdot|$, applied to a diagonal matrix, indicates the entrywise absolute value.
\begin{definition}[Random walk spectral embedding into $\R^{d-1}$]
  \label{def:embedding}
  Given a connected graph and an integer $d$, suppose the eigendecomposition of $\Lrw$ is $\Lrw = \sum_i \hat{\lambda}_i \hat{u}_i \hat{u}_i^\top$ with eigenvalues in the order $|\hat{\lambda}_1|~\geq~\ldots~\geq~|\hat{\lambda}_n|$, and eigenvectors $\hat{u}_1,\ldots,\hat{u}_n$. Let $\hat{\+U} = (\hat{u}_2 , \ldots , \hat{u}_d) \in \R^{n \times (d-1)}$ and $\hat{\+S} = \diag(\hat \lambda_2,\ldots,\hat \lambda_d) \in \R^{(d-1)\times (d-1)}$ and define the random walk spectral embedding as $\hat{\+X} = (\hat{X}_1,\ldots,\hat{X}_n)^\top :=  \hat{\+U}|\hat{\+S}|^{1/2} \in \R^{n \times (d-1)}$.
\end{definition}

We highlight the omission of the first, constant eigenvector $\hat{u}_1 \propto \mathbf{1}$ in the spectral embedding, so that we include only the first $d-1$ non-trivial eigenvectors. It should also be noted that the eigenvectors of $\Lrw$ are not unique and, since it is not a symmetric matrix, the usual choice of an orthonormal system is not available. For our statistical interpretation of Definition~\ref{def:embedding}, which we present in Section~\ref{sec:theory}, we choose $\hat{u}_i = \+D^{-1/2}\breve{u}_i$ as a canonical eigenvector, where $\breve{u}_1,\breve{u}_2,\ldots$ form an orthonormal system of eigenvectors for $\Lsym$. Any other choice (such as unit length eigenvectors) amounts only to coordinate-wise rescaling of the embedding, which in practice is often immaterial because many subsequent inference procedures (such as Gaussian clustering, linear regression) are invariant to such transformations. As a final note, if the graph under study is disconnected, the procedure can be applied to each connected component separately. 

\section{Random graph models}
\label{sec:models}
As a framework for understanding the properties of random walk spectral embedding, we consider a flexible generative model for independent-edge random graphs, known as the generalised random dot product graph \citep{rubin2017statistical}.
\begin{definition}[The generalised random dot product graph]
  Let $F$ be a distribution on a set $\mathcal{X} \subset \R^d$ satisfying $x^\top\Ipq y \in [0,1]$ for all $x,y \in \mathcal{X}$, with full-rank second moment matrix $\E(X X^\top)$ for $X \sim F$. A generalised random dot product graph, with signature $(p,q)$, has latent positions $X_1,\ldots,X_n \overset{\text{i.i.d.}}{\sim} F$, conditional upon which its adjacency matrix $\+A \in \{0,1\}^{n\times n}$ is symmetric and hollow with
  \begin{equation*}
    \+A_{ij} \overset{\text{ind.}}{\sim} \text{Bernoulli}(X_i^\top\Ipq X_j),
  \end{equation*}
  for $i < j$.
\end{definition}

The model is parametrised by a collection of $n$, $d$-dimensional latent positions, one for each node of the graph. It is important to note that the latent positions are not fully identifiable. Replacing each $X_i$ with $\+Q X_i$, where $\+Q$ is in the group $\O(p,q) = \{\+M : \+M^\top\+I_{p,q}\+M = \+I_{p,q}\}$, does not change the conditional distribution of $\+A$. For this reason, our estimation results hold up to some unknown joint transformation $\+Q \in \O(p,q)$.

The generalised random dot product graph contains many popular random graph models as special cases, including the standard and degree-corrected stochastic block models \citep{holland1983stochastic, karrer2011stochastic}.

\begin{definition}[Degree-corrected stochastic block model]
  A graph is said to follow a degree-corrected stochastic block model if its nodes can be partitioned into $K$ communities, conditional upon which,
  \begin{equation*}
    \+A_{ij} \overset{\text{ind.}}{\sim} \text{Bernoulli}(w_i w_j\+B_{z_iz_j}),
  \end{equation*}
  where $\+B \in [0,1]^{K\times K}$, $w_i \in (0,1]$ is a node specific parameter and $z_i \in \{1,\ldots,K\}$ is an index denoting the community membership of the $i$th node.
\end{definition}
If we additionally assume that the communities $z_1,\dots, z_n$ are independently assigned according to some probability vector $\pi$ and that, conditional on this assignment, $w_i \overset{\text{ind.}}{\sim} H_{z_i}$ for some distributions $H_1,\ldots,H_K$ on $(0,1]$, the degree-corrected stochastic block model admits a representation as a generalised random dot product graph. 
Its latent positions are $X_i = w_i v_{z_i}$, where $v_1,\ldots,v_K, p,q$ are such that $\+B_{k \ell}=v_k^\top\Ipq v_{\ell}, k,\ell \in \{1,\ldots,K\}$  with $d = p+q = \text{rank}(\+B) \leq K$. In words, the latent positions belonging to community $k$ live on the one-dimensional subspace spanned by $v_k$, hereafter described as a ``ray'', as shown in Figure~\ref{fig:illustration}a).
The stochastic block model is a special case where $w_1 = \cdots = w_n = 1$ with probability one.

\section{Estimation theory}
\label{sec:theory}

\begin{figure}[t]
  \includegraphics[width=\textwidth]{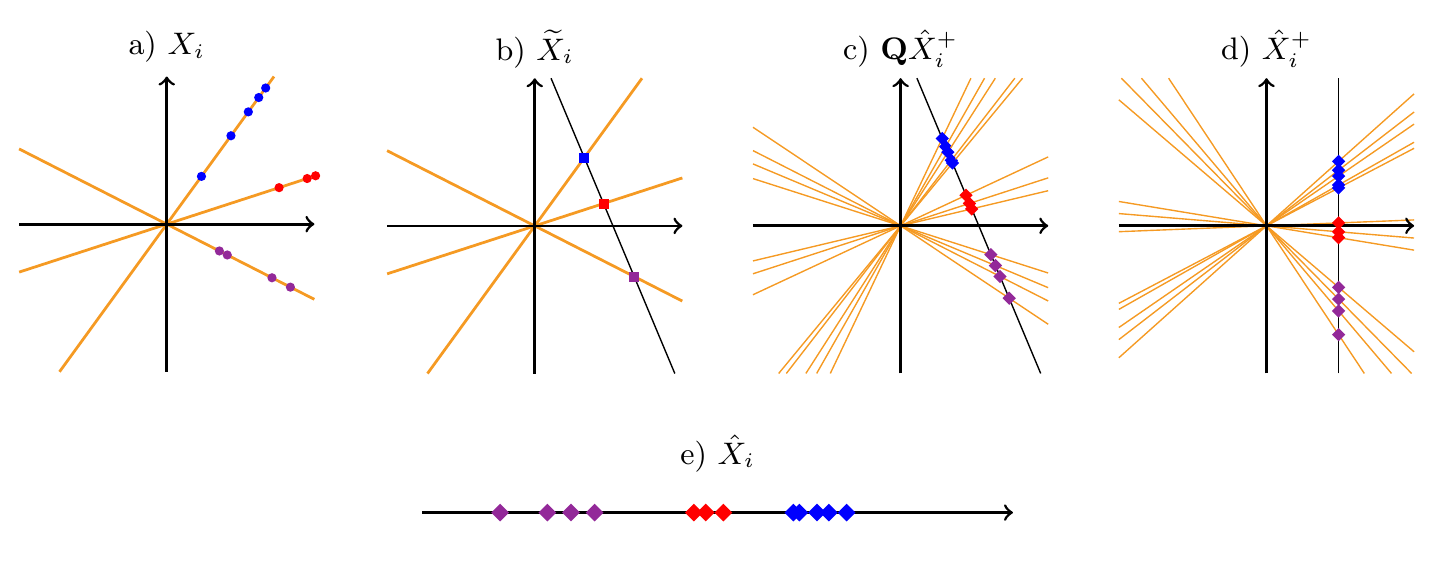}
  % \centering
  % \begin{minipage}{.24\textwidth}
  %   \centering
  %   \include{tikz/X}
  % \end{minipage}
  % \begin{minipage}{.24\textwidth}
  %   \centering
  %   \include{tikz/Xtilde}
  % \end{minipage}
  % \begin{minipage}{.24\textwidth}
  %   \centering
  %   \include{tikz/XhatQ}
  % \end{minipage}
  % \begin{minipage}{.24\textwidth}
  %   \centering
  %   \include{tikz/Xhat}
  % \end{minipage}
  % \begin{minipage}{\textwidth}
  %   \centering
  %   \include{tikz/clt}
  % \end{minipage}
  \caption{Theory pipeline. a) Latent positions in $\R^2$, corresponding to a degree-corrected stochastic block model with three communities; b) Degree-corrected latent positions, which lie on a one-dimensional hyperplane. Latent positions corresponding to the same community have the same degree-corrected latent position; c-d) $d$-dimensional random walk spectral embedding which, in c), is aligned to match the true degree-corrected latent positions; e) $d-1$-dimensional random walk spectral embedding (Definition~\ref{def:embedding}), for input to subsequent clustering step.} %
  \label{fig:illustration} 
\end{figure}

We now make precise the sense in which spectral embedding using the random walk Laplacian produces estimates of the degree-corrected latent positions of a generalised random dot product graph. 

In this paper, a node has an observed degree, $d_i$, and an expected degree, $t_i$; the latter is defined conditionally on the latent positions, $t_i = \sum_j X_i^\top\+I_{p,q} X_j$, and we simply refer to ``degree'' when it is clear which we mean. When we wish to do inference in a way that is agnostic to degree, we want to treat positions differing only by degree as equivalent. Since the degree of a node is proportional to the magnitude of its latent position, this equivalence class is a ray. The space of rays is most conveniently represented through a projective plane, that is, a hyperplane that doesn't pass through the origin. A point on this hyperplane represents the ray that passes through it, and is known as its projective point. If we choose the hyperplane $\{x \in \R^d: (\+I_{p,q} \sum_j X_j)^\top x = 1\}$, the projective point of a ray $\{w X_i: w \in \R\}$ is
\begin{equation*}
  \tilde{X}_i := \frac{X_i}{t_i},
\end{equation*}
which we call a degree-corrected latent position. Figure~\ref{fig:illustration}a) provides an illustration of some latent positions in $\R^2$ (shown as coloured dots) lying on one of three rays (shown as orange lines). Figure~\ref{fig:illustration}b) displays their associated degree-corrected latent positions (shown as coloured squares) and the one-dimensional hyperplane on which they live (shown as a black line). 

We choose to present the theorems that follow in terms of the spectral embedding $\hat{\mathbf{X}}^+ = (\hat{X}^+_1,\ldots,\hat{X}^+_n)^\top := (\hat u_1 , \hat{\+X}) \in \R^{n\times d}$ which includes the trivial canonical eigenvector $\hat u_1 =  \+1/\sqrt{\sum d_i}$ (and implicitly its trivial eigenvalue, 1) which is omitted in Definition~\ref{def:embedding}. This embedding falls entirely on a hyperplane which, again, can be interpreted as a projective plane representing rays. Figure~\ref{fig:illustration}d) provides an illustration of such an embedding (shown as coloured diamonds) and associated rays (shown as orange lines), with the corresponding $d-1$-dimensional embedding $\hat{X}_1,\ldots,\hat{X}_n$ shown in Figure~\ref{fig:illustration}e).

\begin{figure*}[t!]
  \centering
  \includegraphics[width=\textwidth]{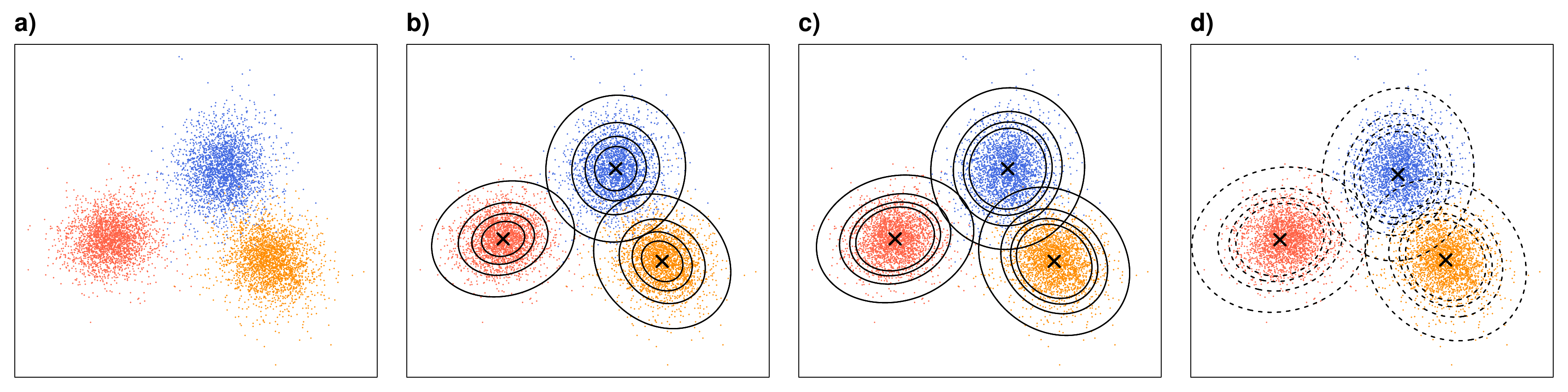}
  \caption{Spectral clustering under a degree-corrected stochastic block model using random walk spectral embedding. a) Spectral embedding of a graph on $n=8000$ nodes, simulated from the degree-corrected stochastic block model described in Eq.~\ref{eq:dcsbm_plot}, coloured according to community membership. b,c) Theoretical means and 95\% level sets of the error distributions, for weights $w_i=0.25,0.5,0.75,1$, for b) dense $(\rho_n = 1)$ and c) sparse $(\rho_n \to 0)$ regimes (after re-alignment and neglecting the first coordinate, see main text for details). d) 95\% level sets of the weighted Gaussian mixture model estimated using the expectation-maximisation algorithm described in Section~\ref{sec:clustering} and the appendix.}
  \label{fig:rwse_embeddings}
\end{figure*}

It is obvious from this diagram that there has been no attempt, in our setup, to make the two projective planes the same, so that the set $\tilde X_1,\ldots, \tilde X_n$ can only resemble  $\hat{X}^+_1, \ldots, \hat{X}^+_n$ after some re-alignment. A matrix $\+Q \in \O(p,q)$ provides the necessary transformation, which can be deterministically constructed given the graph \emph{and} the true (but unidentifiable) latent positions, so is available in simulation (e.g. in Figure~\ref{fig:rwse_embeddings}) but not in practice. Figure~\ref{fig:illustration}c) shows an illustration of the action of $\+Q$.

We consider an asymptotic regime in which we let the number of nodes in the graph grow. In order to facilitate graphs whose expected degree grows more slowly than $n$, we allow the latent positions to shrink as the graph grows. To achieve this, a sparsity factor $\rho_n$ is introduced and, instead of drawing $X_i$ directly from $F$, $\xi_i$ is drawn from $F$ and we set $X_i^{(n)} := \rho_n^{1/2}\xi_i$. Here, $\rho_n$ is a positive sequence such that either $\rho_n=1$ or $\rho_n \to 0$. In the special case of a degree-corrected stochastic block model, the inter-community link probability matrix is $\rho_n \+B$ but we will continue to define $v_k$ in terms of $\+B$, so that $X^{(n)}_i = \rho_n^{1/2} v_{z_i}$. While we talk of the ``graph growing'', technically a new graph is drawn at each $n$, based on $X_1^{(n)}, \ldots, X_n^{(n)}$. We now drop the index $n$ from the notation, with this dependence understood. If $\rho_n \to 0$, we require that it doesn't shrink too quickly, so that the average degree grows at least polylogarithmically, a statement we make precise in the theorems. Finally, we assume $F$ (under the generalised random dot product graph) and $\+B$ (under the degree-corrected stochastic block model) are such that the graph is connected with high probability, a sufficient but by no means necessary condition being that all induced edge probabilities are strictly positive.

\begin{theorem}[Uniform consistency]
  \label{thm:tti}
  Under a generalised random product graph, there exists a universal constant $c \geq 1$ such that, provided the sparsity factor satisfies $n\rho_n = \omega(\log^{4c} n)$,

  \begin{equation}
    \label{eq:tti}
    \max_{i \in \{1,\ldots,n\}} \left\| \+Q \hat{X}^+_i - \tilde{X}_i
     \right \| = O_\P\left(\frac{\log^c n}{n^{3/2}\rho_n}\right).
  \end{equation}
\end{theorem}
Here, a random variable $Y = O_{\P}(f(n))$ if, for any positive constant $\alpha > 0$ there exists an integer $n_0$ and a constant $C > 0$ (both of which possibly depend on $\alpha$) such that for all $n \geq n_0, |Y|\leq C f(n)$ with probability at least $1-n^{-\alpha}$.

We now consider a fixed, finite subset of nodes, indexed without loss of generality as $1,\ldots,m$, and obtain a central limit theorem on the corresponding errors.  

\begin{theorem}[Central limit theorem]
  \label{thm:clt}
Consider the setting of Theorem~\ref{thm:tti}.  Conditional on $X_i$, for $i=1,\ldots,m$, the random vectors $n^{3/2}\rho_n ( \+Q \hat{X}_i^+ - \tilde{X}_i )$
 converge in distribution to independent mean-zero normal random vectors with covariance matrices $\+\Sigma(\rho_n^{-1/2}X_i)$ respectively, where
  \begin{equation*}
    \+\Sigma(x) = \frac{\+I_{p,q}\+\Delta^{-1}\+\Gamma_{\rho}(x) \+\Delta^{-1}\+I_{p,q}}{(\mu^\top\+I_{p,q}x)^2}
  \end{equation*}
with 
\begin{equation*}
  \+\Gamma_{\rho}(x) = 
  \begin{cases}
  \E \left\{ (x^\top \+I_{p,q} \xi)(1 - x^\top \+I_{p,q} \xi) \left( \frac{\xi}{\mu^\top\Ipq \xi} - \frac{\+\Delta\Ipq x}{\mu^\top \Ipq x}  \right)\left( \frac{\xi}{\mu^\top\Ipq \xi} - \frac{\+\Delta\Ipq x}{\mu^\top \Ipq x}  \right)^\top  \right\} & \text{if } \rho_n = 1, \\
  \E \left\{ (x^\top \+I_{p,q} \xi) \left( \frac{\xi}{\mu^\top\Ipq \xi} - \frac{\+\Delta\Ipq x}{\mu^\top \Ipq x}  \right)\left( \frac{\xi}{\mu^\top\Ipq \xi} - \frac{\+\Delta\Ipq x}{\mu^\top \Ipq x}  \right)^\top  \right\} & \text{if } \rho_n \to 0,
  \end{cases}
\end{equation*}
where $\xi \sim F, \mu = \E(\xi)$, $\+\Delta = \E(\frac{\xi\xi^\top}{\mu^\top\Ipq\xi})$.
\end{theorem}
To be clear, Theorem~\ref{thm:clt} is a central limit theorem for a set of $d$-dimensional vectors which, with probability one, live together on a $d-1$-dimensional hyperplane. Accordingly, the derived covariance matrices have rank $d-1$. 

The details of the proofs of Theorems~\ref{thm:tti} and \ref{thm:clt} are given in the appendix. The employed proof mechanism exploits the relationship between the eigenvectors of the random walk Laplacian and the symmetric Laplacian, described in Section~\ref{sec:rwl}, and makes extensive use of the results derived in \citep{rubin2017statistical} and \citep{tang2018limit}.

As a special case of Theorem~\ref{thm:clt}, under the degree-corrected stochastic block model, each estimate is distributed around its community's degree-corrected latent position with asymptotically Gaussian error. 

\begin{corollary}
  \label{cor:dcsbm_clt}  
  Under a degree-corrected stochastic block model with sparsity factor $\rho_n$ satisfying the conditions of Theorem~\ref{thm:tti}, let $\tilde{v}_k = \rho_n^{-1/2} v_{k}/\sum_j w_j \+B_{k z_j}$. Conditional on $z_i$ and $w_i$, for $i = 1, \ldots, m$, the random vectors $n^{3/2}\rho_n ( \+Q \hat{X}_i - \tilde{v}_{z_i} ) $ converge in distribution to independent mean-zero normal random vectors with covariance matrices
  \begin{equation*}
    \+\Sigma(z_i, w_i) =  \frac{\sum_{\ell=1}^K \pi_\ell\+I_{p,q}\+\Delta^{-1}\+\Gamma_{\ell}(z_i, w_i) \+\Delta^{-1}\+I_{p,q}}{w_i\omega_{z_i}^2},
  \end{equation*}
  respectively, with
  \begin{equation*}
  \+\Gamma_{\ell}(k, w) = 
  \begin{cases}
  \E \left( \theta_\ell\+B_{k\ell}(1 - w\theta_\ell\+B_{k\ell})\right) \left( \frac{v_\ell}{\omega_\ell} - \frac{\+\Delta\Ipq v_k}{\omega_k} \right) \left( \frac{v_\ell}{\omega_\ell} - \frac{\+\Delta\Ipq v_k}{\omega_k} \right)^\top  & \text{if } \rho_n = 1, \\
  \E \left( \theta_\ell\+B_{k\ell} \right) \left( \frac{v_\ell}{\omega_\ell} - \frac{\+\Delta\Ipq v_k}{\omega_k} \right) \left( \frac{v_\ell}{\omega_\ell} - \frac{\+\Delta\Ipq v_k}{\omega_k} \right)^\top  & \text{if } \rho_n \to 0,
  \end{cases}
  \end{equation*}
  where $\theta_\ell \sim H_\ell$, $\omega_\ell = \sum_{m=1}^K \pi_m \E(\theta_m) \+B_{\ell m}$, $\+\Delta = \sum_{m=1}^K \frac{\pi_m \E(\theta_m) v_m v_m^\top}{\omega_m}$.
\end{corollary}

Figure \ref{fig:rwse_embeddings}a) shows the spectral embedding $\hat X_1, \ldots, \hat X_n$ of a graph generated from a degree-corrected stochastic block model with $n=8000$ nodes and parameters
\begin{equation}
  \label{eq:dcsbm_plot}
  \+B = \begin{pmatrix}
    0.08 & 0.06 & 0.06 \\
    0.06 & 0.10 & 0.06 \\
    0.06 & 0.06 & 0.12
  \end{pmatrix}, \quad w_1,\ldots,w_n \overset{\text{i.i.d.}}{\sim} \text{Uniform}(0.25,1), \quad \pi = (1/3,1/3,1/3),
\end{equation}
coloured according to community membership. To obtain Figures~\ref{fig:rwse_embeddings}b,c), we first compute $\+Q^{-1}$ to align the degree-corrected community latent positions $\tilde{v}_1,\ldots,\tilde{v}_3$ with $\hat X^+_1, \ldots, \hat X^+_n$. After this transformation, the induced theoretical error distributions have no error in the first coordinate, so we do not display it, showing only what happens in the second and third coordinates. Correspondingly, we remove the first coordinate from $\hat X^+_1, \ldots, \hat X^+_n$, to give the embedding $\hat X_1, \ldots, \hat X_n$. The second and third coordinates of the aligned degree-corrected community latent positions $\+Q^{-1}\tilde{v}_1,\ldots,\+Q^{-1}\tilde{v}_3$ are shown as crosses. Figure~\ref{fig:rwse_embeddings}b) shows four ellipses for each community describing the $95\%$ level sets of the aligned, theoretical error distributions for weights $w_i = 0.25, 0.5, 0.75, 1$, assuming the graph is dense ($\rho_n = 1$). Figure~\ref{fig:rwse_embeddings}c) shows the same assuming the graph is sparse ($\rho_n \to 0$). 

\section{Spectral clustering under the degree-corrected stochastic block model}
\label{sec:spec_clust}
In this section we focus on the methodological implications of the estimation theory in Section~\ref{sec:theory}, suggesting a new spectral clustering algorithm, which we compare to a collection of other popular methods. Most existing methods for spectral clustering under the standard or degree-corrected stochastic block model follow the steps in Algorithm~\ref{alg:spec} \citep{ng2001spectral,von2007tutorial,rohe2011spectral,qin2013regularized,lyzinski2014perfect,lei2015consistency,jin2015fast,sarkar2015role,athreya2016limit,tang2018limit,rubin2017statistical,passino2020spectral}, where options are given in brackets, using the initialisms ASE: Adjacency Spectral Embedding; LSE: symmetric Laplacian Spectral Embedding; RWSE: Random Walk Spectral Embedding; GMM: Gaussian Mixture Modelling; WGMM: Weighted Gaussian Mixture Modelling.

 \begin{algorithm}[H]
 \caption{Fitting a degree-corrected stochastic block model (spectral clustering)}\label{alg:spec}
\begin{algorithmic}[1]
  \Statex \textbf{input} adjacency matrix $\+A$, dimension $d$, number of communities $K\geq d$
  \State Compute spectral embedding $\hat X_1, \ldots, \hat X_n$ of the graph (ASE/LSE/RWSE)
  \State Apply degree-correction (first eigenvector/spherical projection/none)
  \State Apply clustering algorithm ($k$-means/GMM/WGMM)
  \Statex \Return community memberships $\hat z_1, \ldots, \hat z_n$
\end{algorithmic}
 \end{algorithm}

Our theory recommends using the \emph{last} option in every step. We will elaborate on weighted Gaussian mixture modelling in the next subsection and, focussing on random walk spectral embedding, compare this to other choices in step 3. In the following subsection, Section~\ref{sec:comparison}, we will compare our proposal with other combinations of choices across steps 1, 2 and 3, in line with existing literature.

The experimental setup is common to both subsections. We have found that comparisons are sensitive to sparsity, class imbalance, $\+B$ and $n$. In the main document, we will fix $\+B$ to Eq.~\ref{eq:dcsbm_plot} (up to scaling in the dense case), draw each $w_i$ uniformly on $[0.1,1]$, and consider four regimes giving sparse/dense, balanced/imbalanced combinations with growing $n$. We obtain a dense condition by multiplying $\+B$ by 5 and, in this case, we halve the range of $n$. In the balanced condition, we set $\pi = (1/3,1/3,1/3)$, whereas in the imbalanced, we set $\pi = (0.6,0.2,0.2)$. In the appendix, we show the same experiments for other choices of $\+B$. 
\subsection{The choice of clustering algorithm}
\label{sec:clustering}
  \begin{figure}[t]
    \centering
    \includegraphics[width=.85\textwidth]{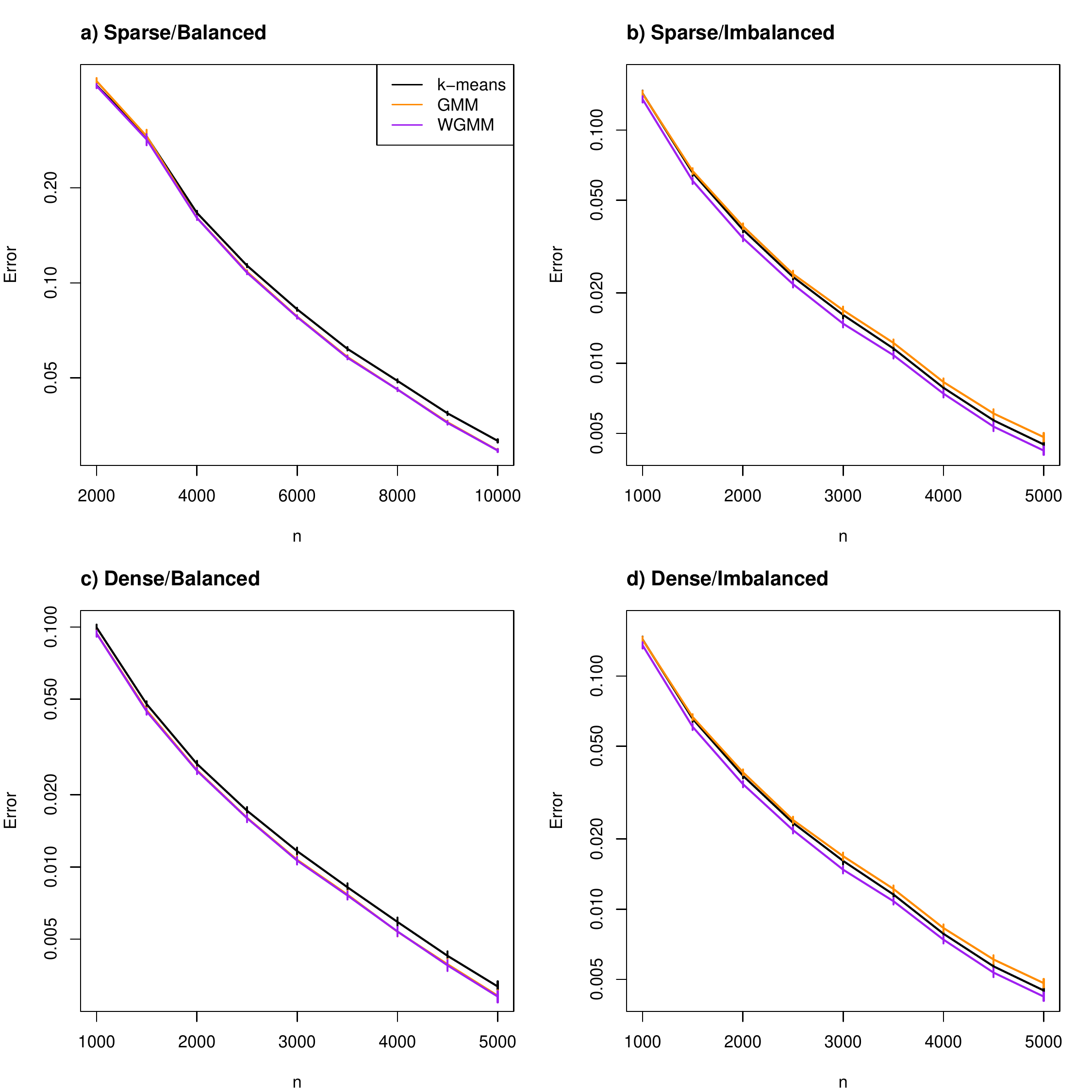}  
    \caption{Comparison of different clustering algorithms applied to the random walk spectral embedding. Graphs are simulated from a degree-corrected stochastic block model (Eq.~\ref{eq:dcsbm_plot}), altered to reflect different regimes. The mean classification error is shown on the log-scale, with the vertical bars showing plus and minus two standard errors, computed over 100 simulated graphs.}
    \label{fig:clustering_comparisons}
  \end{figure}

  Under random walk spectral embedding, the scale of the error covariance matrix about the true degree-corrected latent position is inversely related to its expected degree, as shown in Theorem~\ref{thm:clt}. In sparse graphs (where $\rho_n \to 0$), this relationship is linear and, in particular, under the sparse degree-corrected stochastic block model, the scale of the error distribution is inversely proportional to the weight parameter $w_i$, while its shape, as seen in Figure~\ref{fig:rwse_embeddings}c), depends only on the community. 
  In short --- higher degree nodes are embedded more accurately than lower degree nodes.

  The uniform consistency results of Theorem~\ref{thm:tti} ensure that, under the degree-corrected stochastic block model, asymptotically perfect clustering can be achieved by applying any reasonable clustering algorithm to the spectral embeddings obtained via the random walk Laplacian. Traditionally in the spectral clustering literature, the recommendation has been to use the $k$-means algorithm. However, central limit theorems for adjacency and symmetric Laplacian spectral embedding under the standard stochastic block model have recently motivated fitting a Gaussian mixture model \citep{athreya2016limit,tang2018limit,rubin2017statistical}, the actual asymptotic distribution of the embeddings.

  Here, the theory recommends a slightly more intricate procedure. Our central limit theorem instead suggests that, under a sparse degree-corrected stochastic block model, the data will fit a \emph{weighted} Gaussian mixture model, with likelihood
  \begin{equation}
    L(\alpha, \mu, \+C; \hat X_1, \ldots, \hat X_n, \gamma) = \prod_{i=1}^n \sum_{k=1}^K \alpha_k f(\hat X_i ; \mu_k, \gamma_i^{-1}\+C_k),
    \label{eq:weighted_likelihood}\end{equation} 
  for mixing proportions $\alpha = (\alpha_1,\ldots, \alpha_K)$ subject to  $\alpha_k\geq 0, \sum_{k}\alpha_k = 1$, means $\mu = (\mu_1,\ldots,\mu_K)$ and covariances $\+C = (\+C_1,\ldots,\+C_K)$, where  $f(\cdot;\cdot)$ is the probability density function of a multivariate normal distribution. Here, the weights $\gamma = (\gamma_1,\ldots,\gamma_n)$ are proportional to the unknown $t_i$, which we propose to replace by empirical estimates $\hat \gamma_i \propto d_i$.
\begin{figure}[t]
  \centering
  \includegraphics[width=0.85\textwidth]{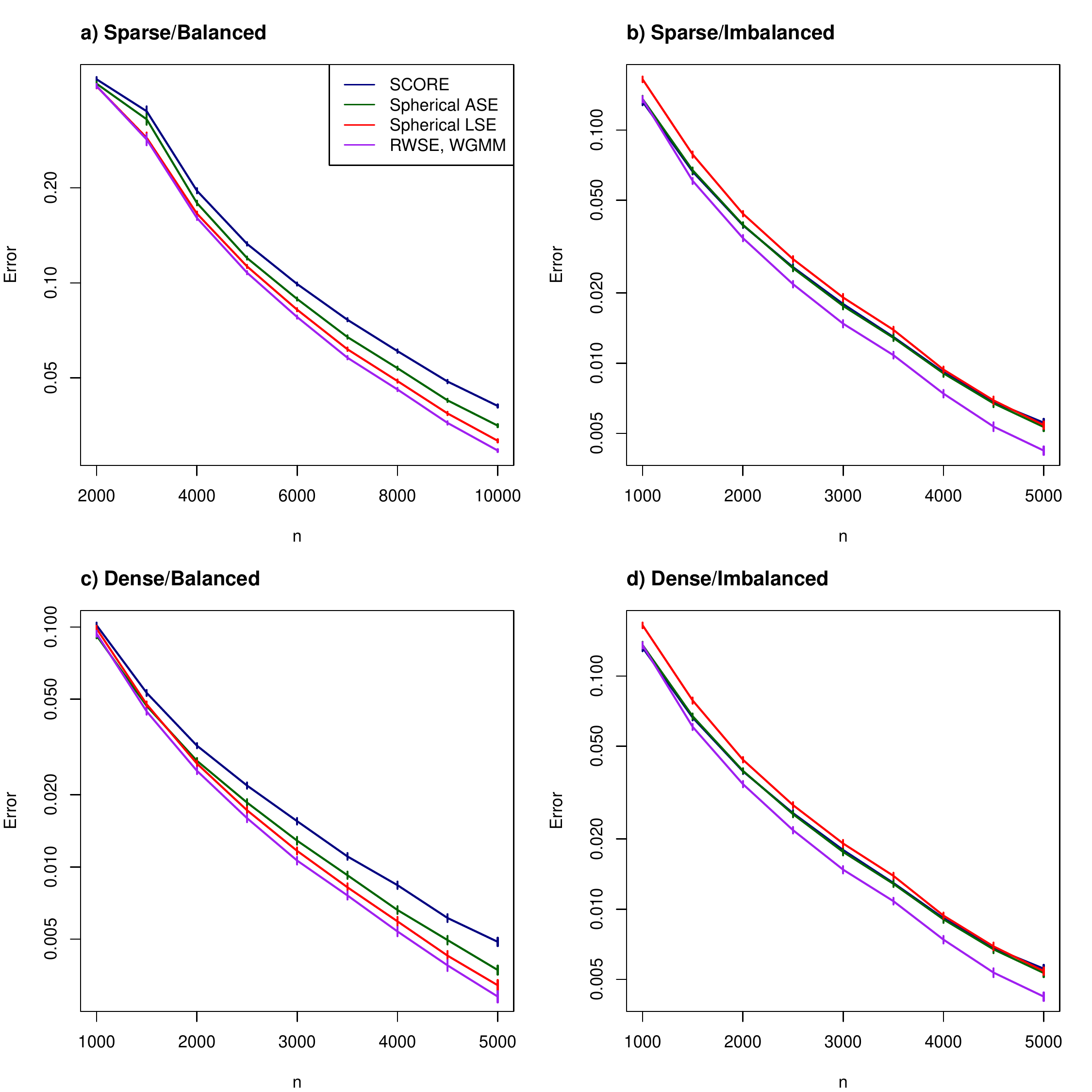}  
  \caption{Comparison of different spectral clustering methods. Graphs are simulated from a degree-corrected stochastic block model (Eq.~\ref{eq:dcsbm_plot}), altered to reflect different regimes. The mean classification error is shown on the log-scale, with the vertical bars showing plus and minus two standard errors, computed over 100 simulated graphs.}
  \label{fig:method_comparisons}
\end{figure}

In the appendix, we provide an expectation-maximisation algorithm to optimise Eq.~\ref{eq:weighted_likelihood}. Upon imposing $\sum_i \hat \gamma_i = n$, this coincides with the expectation-maximisation algorithm for a standard Gaussian mixture model, with the exception that each data point is reweighted according to the parameters $\hat \gamma_1, \ldots, \hat \gamma_n$. Figure~\ref{fig:rwse_embeddings}d) shows the model fit by our algorithm when applied to the spectral embedding $\hat X_1, \ldots, \hat X_n$ obtained from the simulated degree-corrected stochastic block model graph described in Eq.~\ref{eq:dcsbm_plot}.

  We now compare the performance of standard and weighted Gaussian mixture modelling (both using our own implementation), as well as $k$-means (standard R implementation), on random walk spectral embeddings. For this, we simulate graphs from degree-corrected stochastic block models with different sparse/dense, balanced/imbalanced conditions, as described earlier. Performance is quantified via the classification error
  \[\min_{\sigma \in S_3} \frac{1}{n} \sum_{i=1}^n \mathbb{I}\{\sigma(\hat z_i) \neq z_i\},\]
  where $S_3$ denotes the group of all permutations of 3 indices.

  Results are shown in Figure~\ref{fig:clustering_comparisons}. Weighted Gaussian mixture modelling performs best, with $k$-means or standard Gaussian mixture modelling coming second or third depending on the condition. However, it could be argued that the differences between the methods are marginal, and any would make a legitimate choice for the practitioner.

\subsection{Comparison with other methods}
\label{sec:comparison}
In this section, we compare our proposed approach, RWSE followed by WGMM, to other methods in the literature. We implement each of the following methods:
\begin{itemize}
  \item SCORE: ASE, followed by degree-correction using the first eigenvector \citep{jin2015fast};
  \item Spherical ASE: ASE, followed by projection onto the unit sphere \citep{lyzinski2014perfect,lei2015consistency,passino2020spectral};
  \item Spherical LSE: LSE, followed by projection onto the unit sphere \citep{ng2001spectral,qin2013regularized},
\end{itemize}
using the first reference in each case for exact implementation details.

\begin{figure}[t]
  \centering
  \includegraphics[width=0.8\textwidth]{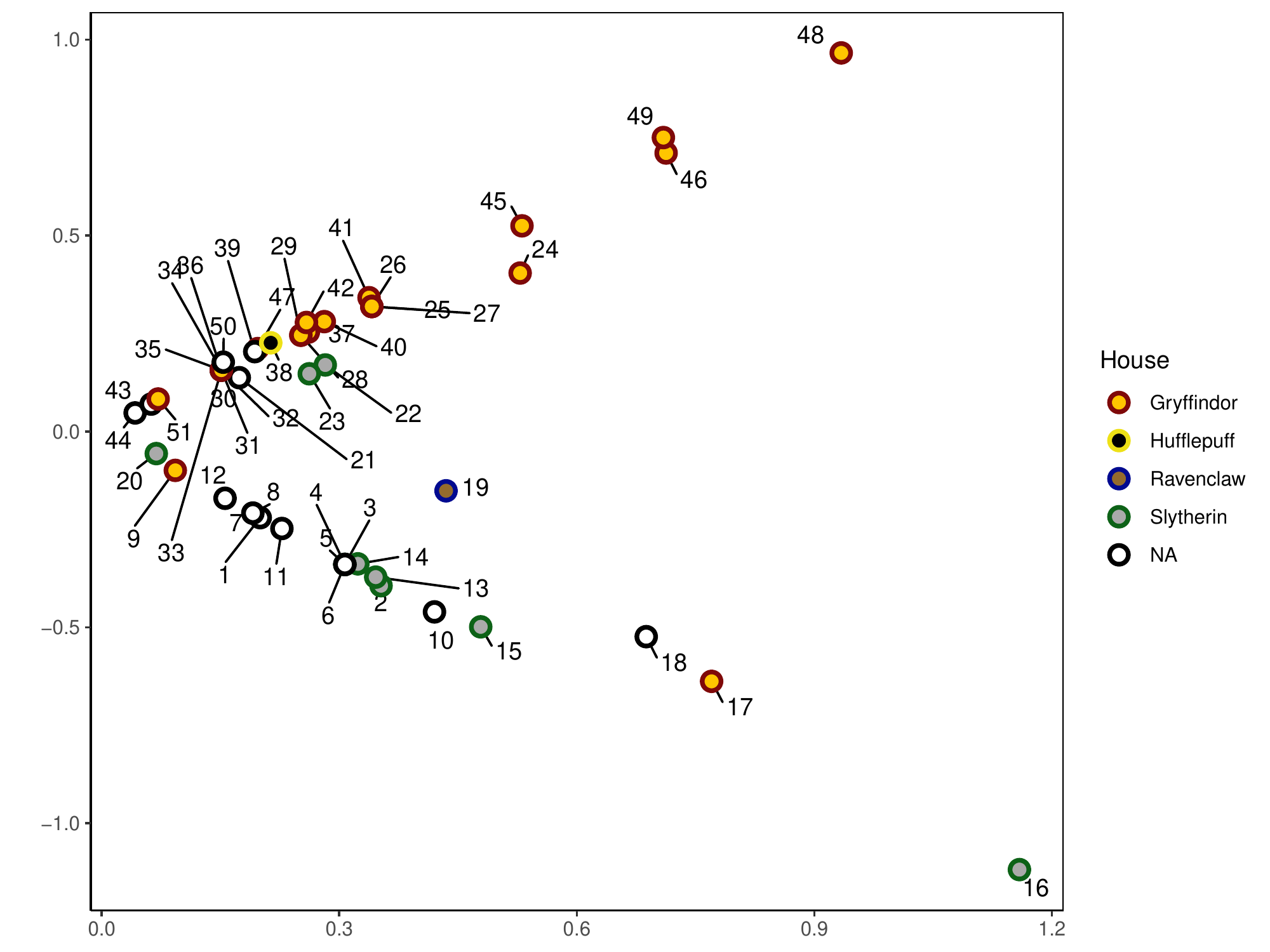}  
  \caption{Two-dimensional adjacency spectral embedding of the Harry Potter enmity network. The numbers correspond to the characters listed in Figure~\ref{fig:hp_rw} and, where applicable, the colours correspond the character's house at Hogwarts.}
  \label{fig:hp_adj}
\end{figure}

As before, we consider four conditions, reflecting sparse/dense and balanced/imbalanced regimes, and report the classification error of each method over 100 Monte Carlo simulations. The results are shown in Figure~\ref{fig:method_comparisons}, with the same for different choices of $\+B$ in the appendix, from which the following conclusions can be drawn. First, across experiments, our method is usually the top performer, especially for large $n$. Second, our method always outperforms spherical LSE. However, in Figures~\ref{fig:method_comparisons2}~and~\ref{fig:method_comparisons3}, our method sometimes performs less well than spherical ASE, especially for low $n$.

Fundamentally, the methods are either based on LSE (spherical LSE and ours) or ASE (spherical ASE and SCORE) and it is known, at a theoretical level, that certain regimes favour one or the other \citep{tang2018limit,cape2019spectral}. In future work, we could investigate how much observed performance differences are driven by this dichotomy.

\section{Real data}
\label{sec:real-data}
In many real world applications, the degree of a node in a network is a parameter of secondary interest. In social networks, we may wish to model a person's friendship preferences independently of their popularity. In cyber-security, and many other domains, the graph represents a snapshot of a dynamic network describing, for example, packet transfers or other network transactions \citep{adams2016dynamic}. The time that a node is present on the network may have a significant bearing on its degree, yet have little to do with its role (e.g. a new laptop connecting to the network). Moreover, the placement of routers and other collection points will result in a higher visibility of some nodes' connections compared to others'. In this case, node degrees are heavily influenced by the observation process and may not be representing an intrinsic property of the nodes themselves.

Stories, real or fictional, often provide network examples to illustrate graph methods, common examples being Zachary's Karate Club \citep{zachary1977information} and the ``Les Miserables'' character network \citep{grover2016node2vec}. Conversely, graph theory is often used in literature studies \citep{labatut2019extraction} to understand character networks and, in this field, degree is often seen as an artifact of the narrative point-of-view: the story spends more time with the protagonist and antagonist, and so we observe more of their connections. As an example, we consider a graph describing the enmity relationships between the characters in the Harry Potter novels of J.K. Rowling \citep{rowling1997}, a publicly available dataset \citep{potterverse2014}. This network has been previously studied in \citep{mara2020csne}.

\begin{figure}[p]
  \centering
  \includegraphics[width=\textwidth]{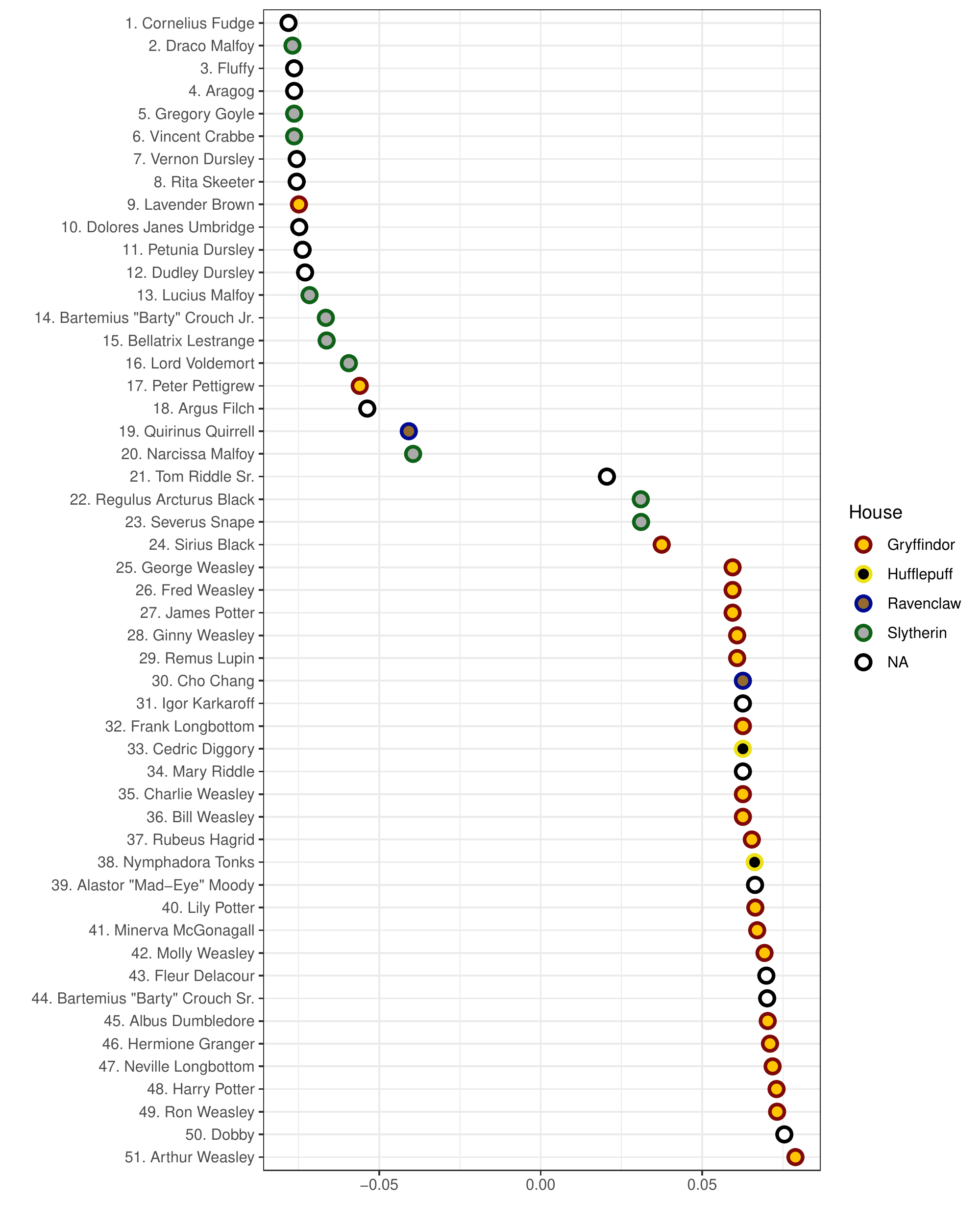}  
  \caption{Random walk spectral embedding of the Harry Potter enmity network into one dimension, coloured, where applicable, according to the character's house at Hogwarts.}
  \label{fig:hp_rw}
\end{figure}

\emph{Those wishing to read the books should refrain from reading the remainder of this section.} Figure~\ref{fig:hp_adj} shows the adjacency spectral embedding of the graph into two dimensions. It shows clear degree heterogeneity among the nodes, with Harry Potter and Voldemort (the protagonist and antagonist of the story, respectively) among the most connected, and minor characters, such as Lavender Brown, among the least. Figure~\ref{fig:hp_rw} shows the random walk spectral embedding of the graph into one dimension, coloured, where possible, according to the characters' house memberships. The embedding shows a clear separation of the characters into two distinct clusters, broadly reflecting their alignment with the protagonist and antagonist. For those unfamiliar with the story, Slytherin tends to house the `evil' characters while Gryffindor tends to house the `good' characters, and this distinction is clearly seen in the positions of each node. However there are exceptions: Severus Snape and Regulus Arcturus Black, members of the Slytherin house, mix in evil circles throughout the story but their benevolence is revealed at late stages in the story, a fact reflected in their positioning. These two outliers are less obvious, in Figure~\ref{fig:hp_adj}, without degree-correction. 

\afterpage{\FloatBarrier}
\newpage
\section{Conclusion}
\label{sec:conclusion}
This paper presents a statistical interpretation of spectral embedding via the random walk Laplacian. Use of the random walk Laplacian, as opposed to the adjacency or symmetric Laplacian matrices, removes the need for post-hoc degree-correction of the spectral embedding, a standard practice in the presence of ancillary degree heterogeneity.

Theoretical results, in the form of uniform consistency and a central limit theorem, support the interpretation that the procedure estimates the degree-corrected latent positions of a generalised random dot product graph. As a result, subsequently applying a standard clustering algorithm, such as $k$-means, fitting a Gaussian mixture model or, better, a weighted Gaussian mixture model, asymptotically achieves perfect clustering under the degree-corrected stochastic block model.

\bibliographystyle{apalike}
\bibliography{random_walk_laplacian}

\appendix

\section{Additional comparisons}
\begin{figure}[p]
  \centering  \includegraphics[width=0.85\textwidth]{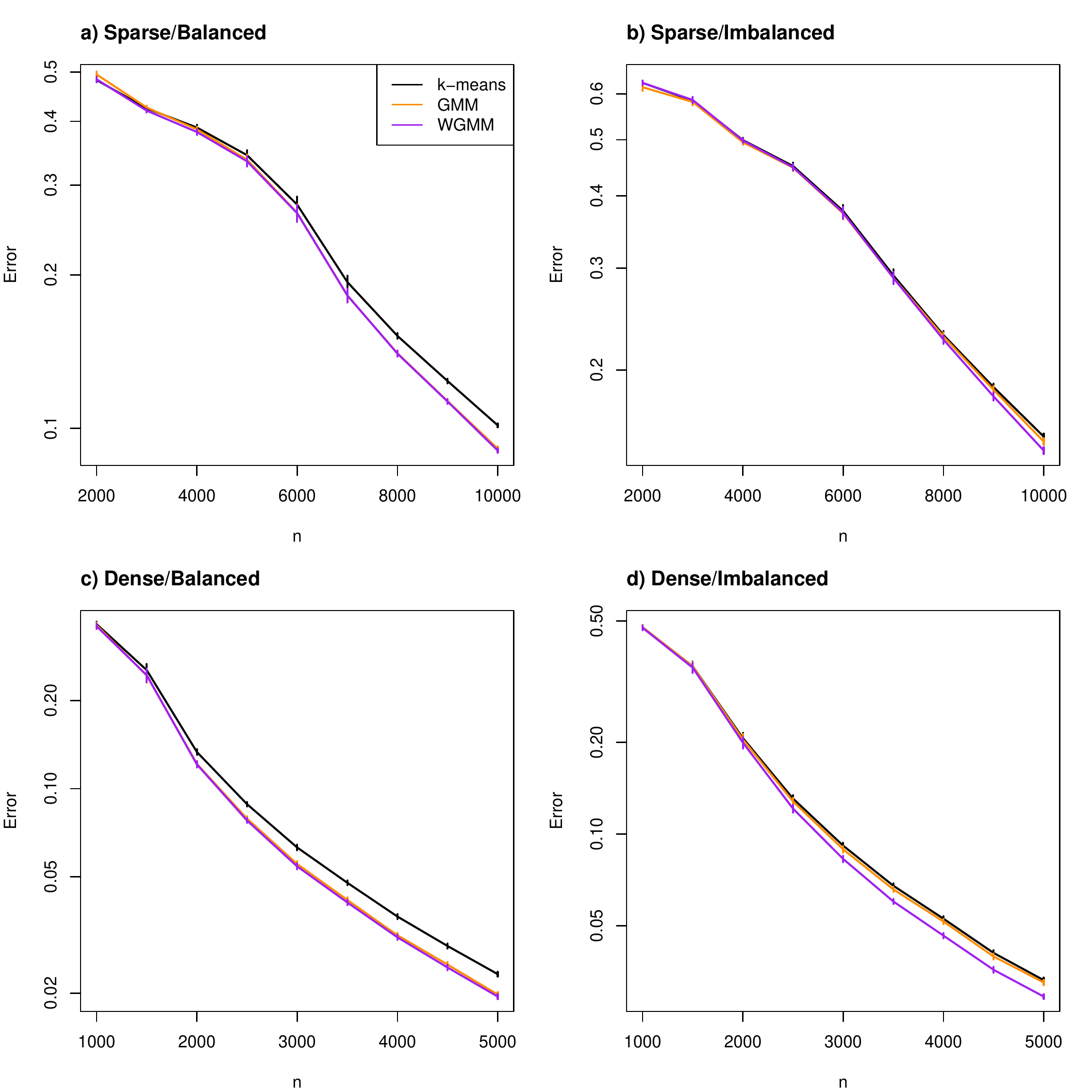}  
  \caption{Comparison of different clustering algorithms applied to the random walk spectral embedding. Graphs are simulated from a degree-corrected stochastic block model (Eq.~\ref{eq:B2}), altered to reflect different regimes. The mean classification error is shown on the log-scale, with the vertical bars showing plus and minus two standard errors, computed over 100 simulated graphs.}
  \label{fig:clustering_comparisons2}
\end{figure}

\begin{figure}[p]
  \centering
  \includegraphics[width=0.85\textwidth]{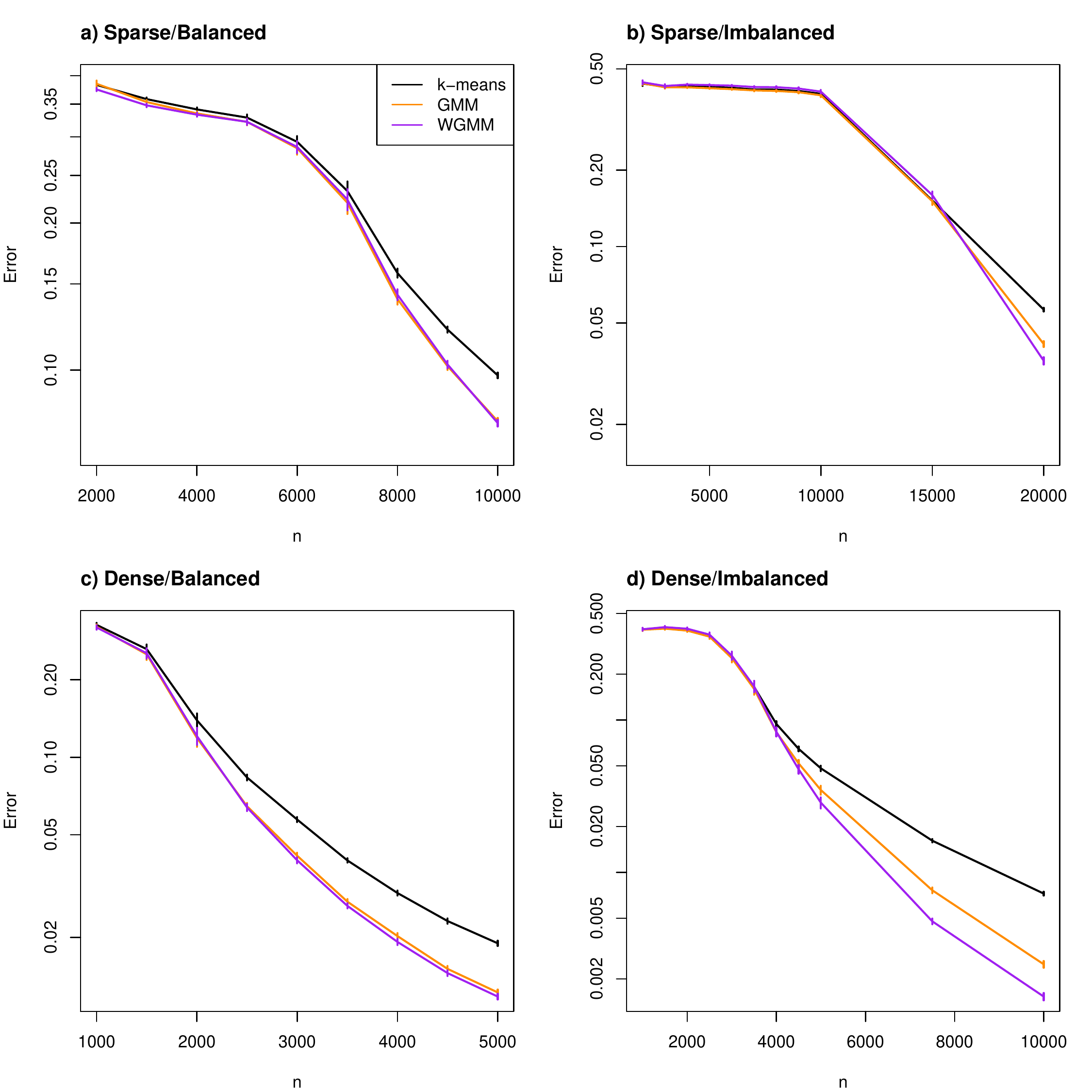}  
  \caption{Comparison of different clustering algorithms applied to the random walk spectral embedding. Graphs are simulated from a degree-corrected stochastic block model (Eq.~\ref{eq:B3}), altered to reflect different regimes. The mean classification error is shown on the log-scale, with the vertical bars showing plus and minus two standard errors, computed over 100 simulated graphs.}
  \label{fig:clustering_comparisons3}
\end{figure}

\begin{figure}[p]
  \centering
  \includegraphics[width=0.85\textwidth]{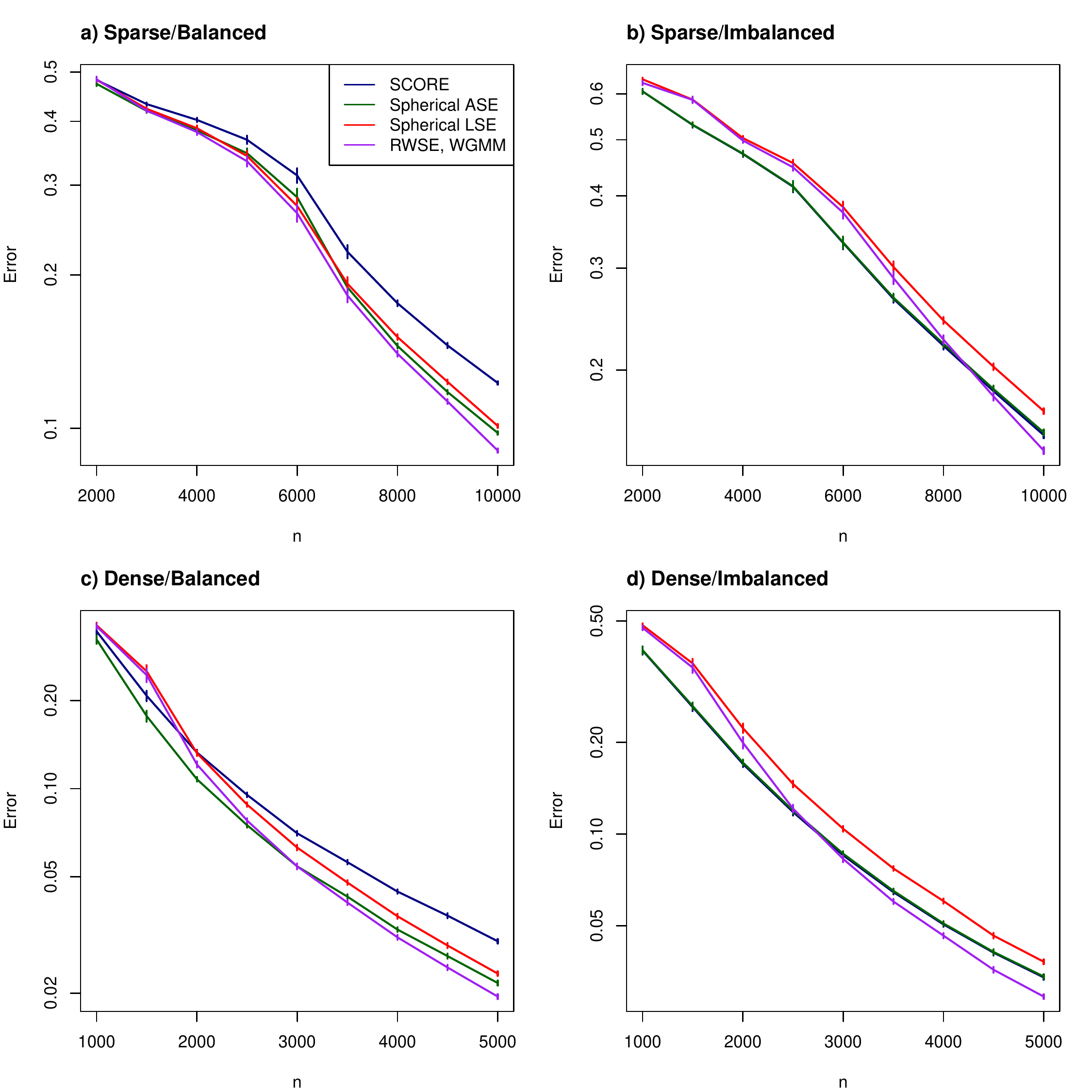}  
  \caption{Comparison of different spectral clustering methods. Graphs are simulated from a degree-corrected stochastic block model (Eq.~\ref{eq:B2}), altered to reflect different regimes. The mean classification error is shown on the log-scale, with the vertical bars showing plus and minus two standard errors, computed over 100 simulated graphs.}
  \label{fig:method_comparisons2}
\end{figure}

\begin{figure}[p]
  \centering \includegraphics[width=0.85\textwidth]{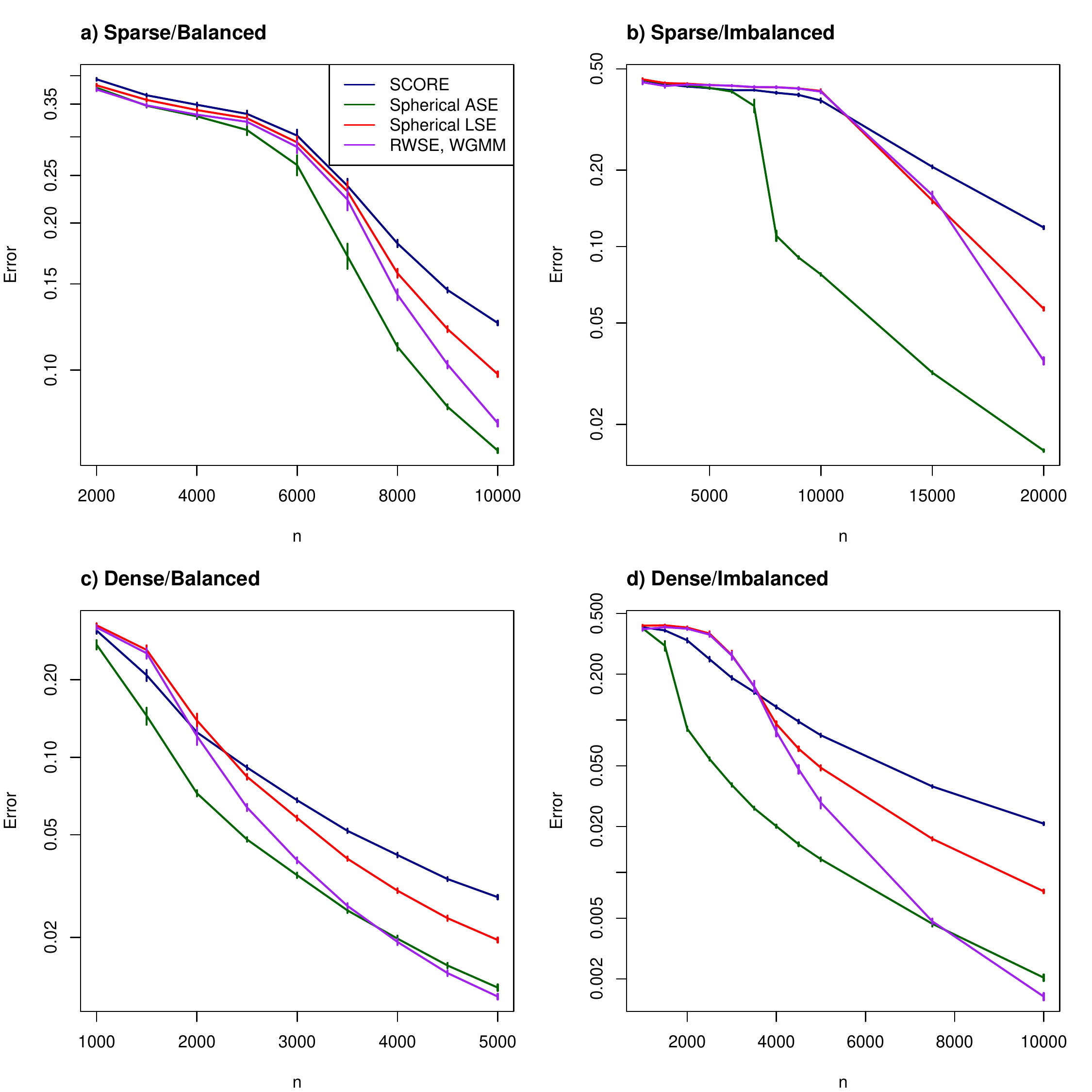} 
  \caption{Comparison of different spectral clustering methods. Graphs are simulated from a degree-corrected stochastic block model (Eq.~\ref{eq:B3}), altered to reflect different regimes. The mean classification error is shown on the log-scale, with the vertical bars showing plus and minus two standard errors, computed over 100 simulated graphs.}
  \label{fig:method_comparisons3}
\end{figure}

The second $\+B$-matrix considered is (divided by 4 in the sparse case)
\begin{equation}
  \+B = \begin{pmatrix}
    0.4 & 0.35 & 0.35 \\
    0.35 & 0.5 & 0.35 \\
    0.35 & 0.35 & 0.6
  \end{pmatrix}
\label{eq:B2}\end{equation}
with results shown in Figures~\ref{fig:clustering_comparisons2}~and~\ref{fig:method_comparisons2}.
The third $\+B$-matrix considered is (divided by 4 in the sparse case)

\begin{equation}
  \+B = \begin{pmatrix}
    0.3 & 0.4 & 0.6 \\
    0.4 & 0.3 & 0.5 \\
    0.6 & 0.5 & 0.3
  \end{pmatrix}
\label{eq:B3}\end{equation}
reflecting disassortative connectivity structure, with results shown in Figures~\ref{fig:clustering_comparisons3}~and~\ref{fig:method_comparisons3}. For both imbalanced cases, we obtained as low an error as we could (as large an $n$ as possible) before reaching memory exhaustion, and could only achieve this by dividing by 4 rather than 5 (which would have matched the experiments in the main text).

\section{Expectation-maximisation algorithm}
Given a set of data points $\hat X_1,\ldots,\hat X_n$ and a set of estimated weights $\hat \gamma_1,\ldots, \hat \gamma_n$, normalised so that $\sum_i \hat \gamma_i =n$, we will optimise Eq.~\ref{eq:weighted_likelihood} for $\hat \mu = (\hat \mu_1,\ldots,\hat \mu_K), \hat{\+C} = ( \hat{\+C}_1\ldots, \hat{\+C}_K)$ as follows.

First, we apply $k$-means to obtain an initial clustering, and set $\hat \beta_{ik} = 1$ if $\hat X_i$ is assigned to the $k$th cluster, and zero otherwise. Then, we alternate between the following steps, starting with the second, until convergence:

\noindent {\bf E-step:} for $i\in\{1,\ldots,n\},k\in\{1,\ldots,K\}$
\begin{align*}
  \hat{\beta}_{ik} &\leftarrow \frac{\hat \alpha_k f(\hat X_i ; \hat{\mu}_k, \hat \gamma_i^{-1}\hat{\+C}_k)}{\sum_\ell \hat{\alpha}_\ell f(\hat X_i ; \hat{\mu}_\ell, \hat \gamma_i^{-1}\hat{\+C}_\ell)}
\end{align*}
\noindent {\bf M-step:} for $k\in\{1,\ldots,K\}$
\begin{align*}
\hat{\alpha}_k &\leftarrow \frac{\sum_{i} \hat{\beta}_{ik}}{n}\\
  \hat{\mu}_k &\leftarrow \left(\frac{1}{\sum_{i} \hat{\beta}_{ik}}\right) \sum_{i=1}^n \hat{\beta}_{ik} \hat \gamma_i \hat X_i\\
  \hat{\+C}_k &\leftarrow \left(\frac{1}{\sum_{i} \hat{\beta}_{ik}}\right) \sum_{i=1}^n \hat{\beta}_{ik} \hat \gamma_i (\hat X_i - \hat{\mu}_k)(\hat X_i - \hat{\mu}_k)^\top
\end{align*}
On convergence, we return the maximum probability membership of each node, $\hat z_i = \text{argmax}_k \hat\beta_{ik}$.

\section{Proofs of Theorems~\ref{thm:tti} and \ref{thm:clt}} 
The proofs of Theorems~\ref{thm:tti} and \ref{thm:clt} make extensive use of the results derived in \citep{rubin2017statistical} and \citep{tang2018limit}. Where the techniques employed here are straight-forward adjustments of those developed in those papers, we refer the reader to the relevant derivations and omit the details.

In what follows, $\norm{\cdot}$ and $\infnorm{\cdot}$ denote the spectral and infinity norms respectively and $\tti{\cdot}$ denotes the two-to-infinity norm \citep{cape2019two}, defined as the maximum row-wise Euclidean norm. We routinely use the inequality $\tti{\+A \+B \+C} \leq \infnorm{\+A} \tti{\+B} \norm{\+C}$, and submultiplicativity of the spectral norm (and, for diagonal matrices, the equivalent infinity norm) without comment.

We define $\+X = (X_1,
\ldots, X_n)^\top \in \R^{n \times d}$, the matrix $\+P = \+X \Ipq  \+X^\top \in \R^{n \times n}$, and for legibility will write $p_{ij}$, $a_{ij}$ to denote $\+P_{ij}$, $\+A_{ij}$ respectively.

Recall that the symmetric Laplacian and random walk Laplacian matrices of $\+A$ are
\begin{equation*}
  \Lsym = \+D^{-1/2}\+A\+D^{-1/2}, \qquad \Lrw = \+D^{-1}\+A
\end{equation*}
where $\+D \in \R^{d \times d}$ is the diagonal degree matrix with entries $d_i = \sum_j a_{ij}$. Suppose that $\breve{\+S}$ is the diagonal matrix containing the $d$ largest-in-magnitude eigenvalues of $\Lsym$ in descending order and $\breve{\+U}$ is the matrix containing corresponding orthonormal eigenvectors as columns. We form analogous objects $\hat{\+S}, \hat{\+U}$ for $\Lrw$. First, let $\hat{\+S}$ be the diagonal matrix of the $d$ largest-in-magnitude eigenvalues of $\Lrw$, observing that $\hat{\+S} = \breve{\+S}$. Second, let $\hat{\+U} := \+D^{-1/2}\breve{\+U}$ contain corresponding canonical eigenvectors.
We define the symmetric Laplacian spectral embedding as $\breve{\+X} = \breve{\+U}|\breve{\+S}|^{1/2}$ and recall that $\hat{\+X}^+ = \hat{\+U}|\hat{\+S}|^{1/2}$. Additionally, we define $\bar{\+X} = \+T^{-1/2}\+X$ where $\+T \in \R^{n \times n}$ is the diagonal expected degree matrix with entries $t_i = \sum_j p_{ij}$ and recall that the degree-corrected latent positions are $\tilde{\+X} := (\tilde X_1,\ldots,\tilde X_n)^\top= \+T^{-1}\+X$ (see Section~\ref{sec:theory}). Since $t_i = \Theta(n\rho_n)$ and $\mathcal{X}$, the support of $F$, is a bounded set \citep{solanki2019persistent}, we have that $\tti{\+X} = O_\P(\rho_n^{1/2})$ and therefore $\tti{\bar{\+X}} = O_\P(n^{-1/2})$. 

A Chernoff bound gives that $|t_i - d_i| = O_\P((n\rho_n)^{1/2}\log n)$ and a union bound gives
\begin{equation}
  \label{eq:T-D}
  \infnorm{\+T - \+D} = O_\P((n\rho_n)^{1/2} \log n).
\end{equation}
Lemma 3.1 of \citep{tang2018limit} gives that $\+D^{-1/2} - \+T^{-1/2}$ admits the decomposition
\begin{equation}
  \label{eq:taylor}
  \+D^{-1/2} - \+T^{-1/2} = \tfrac{1}{2}\+T^{-3/2}(\+T - \+D) + \+R_1
\end{equation}
where $\+R_1$ is a diagonal matrix satisfying $\infnorm{\+R_1} = O_\P((n\rho_n)^{-3/2}\log n)$.

We now reproduce Theorem 7 from \citep{rubin2017statistical} which states that there exists a universal constant $c \geq 1$ and a matrix $\+Q$ such that the symmetric Laplacian spectral embedding satisfies
\begin{equation}
  \label{eq:Lsym_tti}
  \tti{\breve{\+X}\+Q^\top - \bar{\+X}} = O_\P\left( \frac{\log^c n}{n\rho_n^{1/2}} \right).
\end{equation}
Recalling that $\hat{\+X}^+ = \+D^{-1/2}\breve{\+X}$ and $\tilde{\+X} = \+T^{-1/2}\bar{\+X}$, we use Eq. \ref{eq:taylor} to obtain
\begin{equation}
  \label{eq:XhatQt-Xtilde}
  \begin{aligned}
    \hat{\+X}^+\+Q^\top - \tilde{\+X} &= \+D^{-1/2}\breve{\+X}\+Q^\top - \+T^{-1/2}\bar{\+X} \\
    &= (\+T^{-1/2} + \tfrac{1}{2}\+T^{-3/2}(\+T - \+D) + \+R_1)\breve{\+X}\+Q^\top - \+T^{-1/2}\bar{\+X} \\
    &= \+T^{-1/2}(\breve{\+X}\+Q^\top - \bar{\+X}) + (\tfrac{1}{2}\+T^{-3/2}(\+T - \+D)+\+R_1)\breve{\+X}\+Q^\top \\
    &= \+T^{-1/2}(\breve{\+X}\+Q^\top - \bar{\+X}) + \tfrac{1}{2}\+T^{-3/2}(\+T - \+D)\bar{\+X} + \+R_2
  \end{aligned}
\end{equation}
where $\+R_2 = \+R_1 \bar{\+X} + (\tfrac{1}{2}\+T^{-3/2}(\+T - \+D) + \+R_1)(\breve{\+X}\+Q^\top - \bar{\+X})$. Eqs.~\ref{eq:T-D} - \ref{eq:Lsym_tti} give that
\begin{equation}
  \label{eq:R2}
  \begin{aligned}
    \tti{\+R_2} &\leq \infnorm{\+R_1} \tti{\bar{\+X}} + \left( \tfrac{1}{2}\infnorm{\+T^{-3/2}}\infnorm{\+T - \+D} + \infnorm{\+R_1} \right)\tti{\breve{\+X}\+Q^\top - \bar{\+X}} \\
    &= O_\P\left( \frac{\log^c n}{n^2\rho_n^{3/2}} \right).
  \end{aligned}
\end{equation}
Therefore,
\begin{align*}
  \tti{\hat{\+X}^+\+Q^\top - \tilde{\+X}} &\leq \infnorm{\+T^{-1/2}}\tti{\breve{\+X}\+Q^\top - \bar{\+X}} + \tfrac{1}{2}\infnorm{\+T^{-3/2}}\infnorm{\+T-\+D}\tti{\bar{\+X}} + \tti{\+R_2} \\
  &= O_\P \left(\frac{\log^c n}{n^{3/2}\rho_n}\right),
\end{align*}
establishing Theorem \ref{thm:tti}. To establish Theorem~\ref{thm:clt}, we first state an important decomposition derived in \citep{tang2018limit} for the symmetric Laplacian spectral embedding. We state the decomposition with a minor modification to accommodate both positive and negative leading eigenvalues, where only positive leading eigenvalues are considered in \citep{tang2018limit} (see \citep{rubin2017statistical}). We have
\begin{equation}
  \label{eq:Lsym_decomp}
  \breve{\+X}\+Q^\top - \bar{\+X} = \+T^{-1/2}(\+A - \+P)\+T^{-1/2}\bar{\+X}(\bar{\+X}^\top\bar{\+X})^{-1}\Ipq + \tfrac{1}{2}\+T^{-1}(\+T - \+D)\bar{\+X} + \+R_3
\end{equation}
where $r_i^{(3)}$, the $i$th row of $\+R_3$, is such that $n\rho_n^{1/2} r_i^{(3)} \overset{\text{p}}{\to} 0$, where $\overset{\text{p}}{\to}$ denotes convergence in probability. Substituting Eq.~\ref{eq:Lsym_decomp} into Eq.~\ref{eq:XhatQt-Xtilde} gives
\begin{align*}
  \hat{\+X}^+\+Q^\top - \tilde{\+X} &= \+T^{-1/2}(\breve{\+X}\+Q^\top - \bar{\+X}) + \tfrac{1}{2}\+T^{-3/2}(\+T - \+D)\bar{\+X} + \+R_2 \\
  &= \+T^{-1/2}\{\+T^{-1/2}(\+A - \+P)\+T^{-1/2}\bar{\+X}(\bar{\+X}^\top\bar{\+X})^{-1}\Ipq \\&\hspace{2cm}+\tfrac{1}{2}\+T^{-1}(\+T - \+D)\bar{\+X} + \+R_3\} + \tfrac{1}{2}\+T^{-3/2}(\+T - \+D)\bar{\+X} + \+R_2 \\
  &= \+T^{-1}(\+A - \+P)\+T^{-1/2}\bar{\+X}(\bar{\+X}^\top\bar{\+X})^{-1}\Ipq + \+T^{-3/2}(\+T-\+D)\bar{\+X} + \+T^{-1/2}\+R_3 + \+R_2\\
  &= \+T^{-1}(\+A - \+P)\+T^{-1}\+X(\+X^\top\+T^{-1}\+X)^{-1}\Ipq + \+T^{-2}(\+T - \+D)\+X + \+R
\end{align*}
where $\+R = \+T^{-1/2}\+R_3 + \+R_2$. From \citep{tang2018limit}, $(\+X^\top\+T^{-1}\+X)^{-1} \to \+\Delta^{-1}$, $t_i/(n\rho_n) \to \xi_i^\top \Ipq \mu$ and  $n\rho_n/t_i \to (\xi^\top\Ipq\mu)^{-1}$ almost surely, the latter by the continuous mapping theorem. Let $r_i, r_i^{(2)}$ denote the $i$th rows of $\+R, \+R_2$ respectively. Again, by the continuous mapping theorem, $(n \rho_n / t_i)^{-1/2}$ tends to a constant almost surely and $n\rho_n^{1/2} r_i^{(3)} \overset{\text{p}}{\to} 0$, so we have that $n^{3/2}\rho_n t_i^{-1/2}r^{(3)}_i \overset{\text{p}}{\to} 0$. By Eq.~\ref{eq:R2}, $n^{3/2}\rho_n r_i^{(2)} \overset{\text{p}}{\to} 0 $ and therefore $n^{3/2}\rho_n r_i \overset{\text{p}}{\to} 0$.
Denote by $\zeta_i$ the $i$th row of $n^{3/2}\rho_n(\hat{\+X}^+\+Q^\top - \tilde{\+X})$. From here on, we use $r$ to denote any random vector such that $r \overset{\text{p}}{\to} 0$, which may change from line to line. We have
\begin{align*}
  \zeta_i &= \Ipq (\+X^\top\+T^{-1}\+X)^{-1} \frac{n^{3/2} \rho_n}{t_i} \left( \sum_j \frac{a_{ij} - p_{ij}}{t_j}X_j \right) + \frac{n^{3/2}\rho_n}{t_i^2} (t_i - d_i) X_i + r \\
  &= \Ipq (\+X^\top\+T^{-1}\+X)^{-1} \frac{(n \rho_n)^{3/2}}{t_i} \left( \sum_j \frac{a_{ij} - p_{ij}}{t_j}\xi_j \right) + \frac{(n\rho_n)^{3/2}}{t_i^2} \sum_j (a_{ij} - p_{ij})\xi_i + r \\
  &= \Ipq (\+X^\top\+T^{-1}\+X)^{-1} \left( \frac{n\rho_n}{t_i} \right) \left( \sum_j \frac{(n\rho_n)^{1/2}(a_{ij} - p_{ij})}{t_j} \xi_j \right) + \left(\frac{n\rho_n}{t_i}\right)^2 \left( \sum_j \frac{(a_{ij} - p_{ij})}{(n\rho_n)^{1/2}} \xi_i \right) + r
\end{align*}
Additionally, by an identical argument to that used to obtain Eq.~(B.3) and (B.4) in \citep{tang2018limit},
\begin{equation*}
   \sum_j \frac{(n\rho_n)^{1/2}(a_{ij} - p_{ij})}{t_j} \xi_j = \sum_j \frac{(a_{ij} - p_{ij})}{(n\rho_n)^{1/2}} \frac{\xi_j}{\xi_j^\top\Ipq\mu} + r
\end{equation*}
and
\begin{equation*}
  \frac{n\rho_n}{t_i}\sum_j \frac{(a_{ij} - p_{ij})}{(n\rho_n)^{1/2}} \xi_i = \Ipq (\+X^\top\+T^{-1}\+X)^{-1} \sum_j \frac{(a_{ij} - p_{ij})}{(n\rho_n)^{1/2}} \frac{\+\Delta\Ipq \xi_i}{\xi_i^\top \Ipq \mu} + r.
\end{equation*}
Therefore
\begin{equation*}
  \label{eq:zeta_i}
  \zeta_i = \frac{n\rho_n}{t_i} \Ipq (\+X^\top\+T^{-1}\+X)^{-1}\sum_j \frac{(a_{ij} - p_{ij})}{(n\rho_n)^{1/2}} \left( \frac{\xi_j}{\xi_j^\top \Ipq \mu} - \frac{\+\Delta \Ipq \xi_i}{\xi_i^\top \Ipq \mu} \right) + r.
\end{equation*}
Conditional on $\xi_i = x_i$,
\begin{equation*}
  \sum_j \frac{(a_{ij} - p_{ij})}{(n\rho_n)^{1/2}} \left( \frac{\xi_j}{\xi_j^\top \Ipq \mu} - \frac{\+\Delta\Ipq\xi_i}{\xi_i^\top \Ipq\mu} \right)
\end{equation*}
is $n^{-1/2}$ times the sum of independent, identically-distributed vectors, each with mean zero and covariance
\begin{equation*}
  \+\Gamma_\rho(x) = \E \left\{ (x^\top \Ipq \xi)(1 - \rho_nx^\top \Ipq \xi) \left( \frac{\xi}{\xi^\top\Ipq\mu} - \frac{\+\Delta \Ipq x}{x^\top\Ipq \mu} \right) \left( \frac{\xi}{\xi^\top\Ipq\mu} - \frac{\+\Delta \Ipq x}{x^\top\Ipq \mu} \right)^\top \right\},
\end{equation*}
so that, by the multivariate central limit theorem, the sum is Gaussian with mean zero and that covariance (ignoring the vanishing contribution of the $i$th vector).
By application of Slutsky's theorem,
\begin{equation*}
  \zeta_i \to \mathcal{N}\left(0, \frac{\Ipq\+\Delta\+\Gamma_\rho(x)\+\Delta\Ipq}{(x^\top\Ipq\mu)^2}\right)
\end{equation*}
and invoking the Cram\'er-Wold device establishes Theorem~\ref{thm:clt}.

\end{document}